\documentclass[twocolumn]{aastex63}
\hypersetup{linkcolor=black,citecolor=cyan,filecolor=black,urlcolor=magenta}
\usepackage{afterpage}

\received{2019 December 17}
\revised{2020 February 20}
\accepted{2020 February 27}

\shorttitle{Quiescent galaxies at z$\sim$3}
\shortauthors{D'Eugenio et al.}

\begin{document}

\title{The Typical Massive Quiescent Galaxy at z$\sim$3 is a Post-starburst}

\correspondingauthor{C. D'Eugenio}
\email{chiara.deugenio@cea.fr}

\author{C. D'Eugenio}
\affiliation{CEA, IRFU, DAp, AIM, Universit\'e Paris-Saclay, Universit\'e Paris Diderot, Sorbonne Paris Cit\'e, CNRS, F-91191 Gif-sur-Yvette, France}

\author{E. Daddi}
\affiliation{CEA, IRFU, DAp, AIM, Universit\'e Paris-Saclay, Universit\'e Paris Diderot, Sorbonne Paris Cit\'e, CNRS, F-91191 Gif-sur-Yvette, France}

\author{R. Gobat}
\affiliation{Instituto de F\'isica, Pontificia Universidad Cat\'olica de Valpara\'iso, Casilla 4059, Valpara\'iso, Chile}

\author{V. Strazzullo}
\affiliation{Faculty of Physics, Ludwig-Maximilians-Universit\"at, Scheinerstr. 1, D--81679 Munich, Germany}

\author{P. Lustig}
\affiliation{Faculty of Physics, Ludwig-Maximilians-Universit\"at, Scheinerstr. 1, D--81679 Munich, Germany}

\author{I. Delvecchio }
\affiliation{CEA, IRFU, DAp, AIM, Universit\'e Paris-Saclay, Universit\'e Paris Diderot, 
Sorbonne Paris Cit\'e, CNRS, F-91191 Gif-sur-Yvette, France}
\affiliation{INAF -- Osservatorio Astronomico di Brera, Via Brera 28, I-20121, Milano, Italy}

\author{S. Jin}
\affiliation{Instituto de Astrofísica de Canarias (IAC), E-38205 La Laguna, Tenerife, Spain}
\affiliation{Universidad de La Laguna, Dpto. Astrofísica, E-38206 La Laguna, Tenerife, Spain}

\author{A. Puglisi}
\affiliation{Center for Extragalactic Astronomy, Durham University, South Road, Durham DH1 3LE, UKe}
\affiliation{CEA, IRFU, DAp, AIM, Universit\'e Paris-Saclay, Universit\'e Paris Diderot, 
Sorbonne Paris Cit\'e, CNRS, F-91191 Gif-sur-Yvette, France}
\affiliation{INAF-Osservatorio Astronomico di Padova, Vicolo dell'Osservatorio, 5, I-35122 Padova, Italy}

\author{A. Calabr\'o}
\affiliation{INAF-Osservatorio Astronomico di Roma, via di Frascati 33, I-00078, Monte Porzio Catone, Italy}
\author{C. Mancini}
\affiliation{Department of Physics and Astronomy, University of Padova, Vicolo dell’Osservatorio, 3, I-35122, Padova, Italy}

\author{M. Dickinson}
\affiliation{National Optical Astronomy Observatories, 950 N Cherry Avenue, Tucson, AZ 85719, USA}

\author{A. Cimatti}
\affiliation{Università di Bologna, Dipartimento di Fisica e Astronomia, Via P. Gobetti 93/2, I-40129, Bologna, Italy}

\author{M. Onodera}
\affiliation{Subaru Telescope, National Astronomical Observatory of Japan, National Institutes of Natural Sciences (NINS), 650 North A'ohoku Place, Hilo, HI 96720, USA}
\affiliation{Department of Astronomical Science, SOKENDAI (The Graduate University for Advanced Studies), 650 North A'ohoku Place, Hilo, HI 
96720, USA}

\begin{abstract}

We have obtained spectroscopic confirmation with \textit{Hubble Space Telescope}  WFC3/G141 of a first sizeable sample of nine quiescent galaxies at 2.4$<$z$<$3.3.
Their average near-UV/optical rest-frame spectrum  is characterized by low attenuation (Av$\sim$0.6 mag) and a strong Balmer break, larger than the 4000 \AA\ break, corresponding to a fairly young age of $\sim$300 Myr. This formally classifies a substantial fraction of classically selected quiescent galaxies at $z\sim3$ as post-starbursts, marking their convergence to the quenching epoch. The rapid spectral evolution with respect to $z\sim1.5$ quiescent galaxies is not matched by an increase of residual star-formation, as judged from the weak detection of [O II]$\lambda$3727 emission, pointing to a flattening of the steep increase in gas fractions previously seen from $z\sim0$ to 1.8. However, radio 3GHz stacked emission implies either much stronger dust-obscured star formation or substantial further evolution in radio-mode AGN activity with respect to $z\sim1.5$.

\end{abstract}


\section{Introduction} \label{sec:intro}
Timing the formation and quenching of the most massive (M$_{\star}>$10$^{11}M_{\odot}$) passively evolving galaxies (PEGs) is subject of intense debate. Several studies kept unveiling their existence at progressively increasing lookback times, placing their formation at z$>$3-4~\citep{gob12, glaze17, Valentino19, forrest19}. Reproducing the observed number density of such galaxies at all epochs is a compelling concern of current galaxy evolution models in cosmological simulations, as the relative importance of the quenching mechanisms is not yet clear~\citep{manbelli}. 
To this end, several photometric samples of high-\textit{z} PEGs exist \citep{straat14, merlin18}, yet spectroscopic confirmation is still needed to precisely assess their redshifts and the degree of contamination by star-forming interlopers, especially at epochs when PEGs might be just starting to emerge. At z$\sim$3 the chances of collecting statistically meaningful samples of PEG spectra are hampered by their rarity and lack of prominent emission lines. Therefore, the assembly of sizable samples of quiescent spectra is essential to improve upon the photometric age/dust-attenuation estimation for the bulk of their stellar populations, to draw relative comparisons of their spectral properties with respect to lower redshift massive PEGs and to reveal any residual emission line that might be linked to star-formation. The latter point provides useful information on the availability of unstable, cold molecular gas which is, itself, a poorly constrained quantity at this epoch.
In this Letter we report on the average stellar population properties in a sizable sample of nine spectroscopically confirmed 2.4$<$z$<$3.3 quiescent galaxies. We constrain their global residual star formation from the average [O II] emission, showing implications on their average gas fraction. We assume a $\Lambda$CDM cosmology with H$_{0}=70$ km s$^{-1}$ Mpc$^{-1}$, $\Omega_{M}=0.27$, $\Omega_{\Lambda}=0.73$ and a Salpeter initial mass function (IMF). Magnitudes are given in the AB photometric system.

\section{Sample selection and analysis}
We selected galaxies from a parent sample of more than 50 reliable PEGs candidates with K$_{tot}<$22.5 in the \cite{mccracken10} catalog with photometric redshifts between 2.5$\leqslant$ z$_{phot}$ $\leqslant$3.5 in the COSMOS field. A first selection was done using the BzK criterion \citep{daddi04} selecting passive BzK galaxies plus objects formally classified as star-forming BzK but having a signal-to-noise ratio S/N$<$5 in B and z bands, to retain passive galaxies as they become fainter in such bands with increasing redshift and decreasing mass. We then selected from this sample \textit{UVJ} quiescent galaxies \citep{williams09}, exploiting photometric redshifts calibrated for high-\textit{z} PEGs  \citep{ono12, strazz15}.
To minimize contamination from dusty star-forming galaxies, objects in \cite{mccracken10} with Spitzer/MIPS 24$\mu$m S/N$>$4 were discarded, except for galaxies with high-confidence passive spectral energy distributions (SEDs). 
Objects having SED fits to optical--NIR broadband photometry consistent with dusty star-forming solutions and contaminated photometry in all images were also discarded.
Among the most massive bona fide passive candidates, 10 galaxies were targeted for \textit{Hubble Space Telescope (HST)} WFC3/IR G141 near-IR\footnote{Rest-frame near-UV/optical.} observations: 9 with $H_{AB}<22$ ($M_{\star}> 1.1\times 10^{11} M_{\odot}$) plus 1 robust candidate with H$_{AB}$=22.9 ($M_{\star}=8\times 10^{10} M_{\odot}$) among the highest-z objects. Scheduling from 1 to 3 orbits (17 in total) according to each source's flux provided low resolution spectra (R=130) with a mean S/N $\sim$4.5 per resolution element for each target and ancillary F160W imaging (HST GO 15229 program). The data were reduced and decontaminated from neighboring sources' overlapping spectra by means of the \texttt{grizli} software package \citep{grizli}\footnote{https://github.com/gbrammer/grizli}. For each target a 2D spectrum was produced and drizzled to a scale of 0.8 times the native pixel size. The 1D spectra were then optimally extracted \citep{horne86} and fitted from 11000 to 16900 \AA\ with \cite{bc03} (BC03) composite stellar populations derived using different star formation histories (SFHs): allowing for a constant, exponentially declining, delayed exponentially declining, and a truncated SFH, where the star formation rate (SFR) drops to 0 after a timescale between $\tau=0.001-1.0$ Gyr. Ages were allowed to vary between t=100 Myr and 10 Gyr, from redshift 5 to 0. The \cite{cal2000} extinction law was adopted with $A_{V}$ varying between 0 and 8 mag. BC03 templates were broadened to the grism resolution using a Gaussian kernel whose FWHM was derived via a Gaussian fit of the 2D light profile of each source.\\   
The broadband photometry  \citep{laigle16} for each galaxy was also fitted imposing the grism-derived redshifts. The addition of the photometric information mostly leads to narrowing down the range of dust extinction values at any confidence level. The detailed analysis  of individual galaxies will be presented in a forthcoming paper. 

\subsection{Quiescence of individual targets} 
We tested the quiescent/dusty star-forming nature of each galaxy adding the $\chi^{2}$ matrices of fits to spectra and photometry and comparing the goodness-of-fit of the best-fitting constant star-forming (CSF) template with that of a solution defined as passive by constraining the best fitting SFH as follows: t$_{50}\geq$0.3 Gyr, A$_{V}<$0.8 mag and t$_{50}/\tau \geq$10, where t$_{50}$ is the lookback time at which the galaxy formed 50\% of its stellar mass. To reject a solution (i.e., to classify it as inconsistent with respect to the other one \textit{and} with the data) its probability had to be $<$0.01, as inferred from their $\chi^{2}$ difference. Star-forming solutions were rejected for all sources except for one, whose redshift was not properly constrained by our data. If this galaxy were in our selected range in redshift, its quiescence would be rejected with 80\% confidence, hence we excluded it. We were thus left with 9 galaxies classified as passive with a median redshift of z$_{m}$=2.808 and a median stellar mass of $<M_{\star}>$=1.8$\times10^{11}$ M$_{\sun}$.

\section{Stacked Spectrum}
\begin{figure}
\centering
\includegraphics[width=\columnwidth]{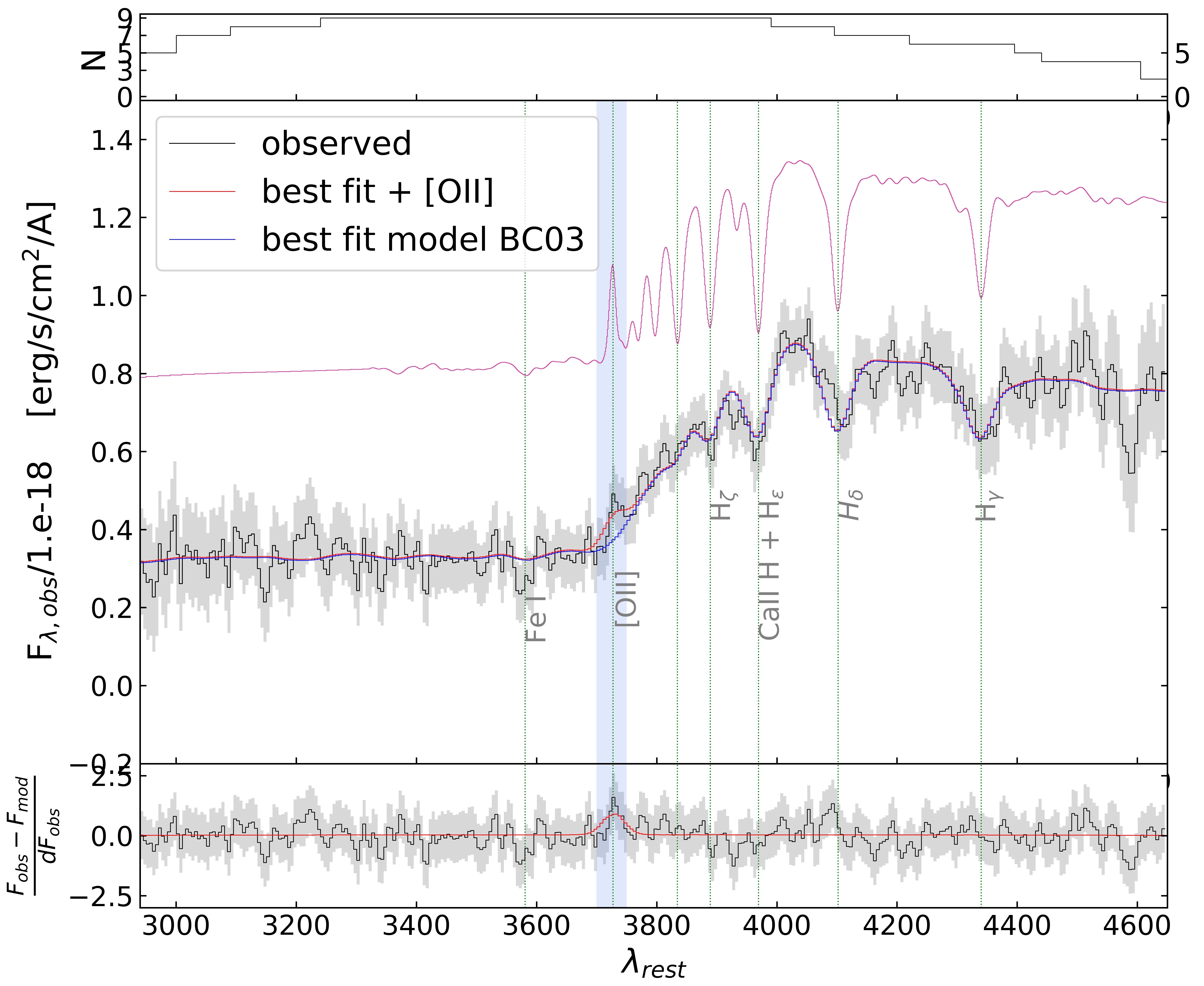}
  \caption{Top: Number of galaxies contributing to each spectral bin. Middle: Stack of 9 quiescent galaxies' spectra. Blue curve: best-fit model from the BC03 library. Red curve: Gaussian fit to the residual, added to the blue continuum. Pink curve: best-fit BC03 template smoothed for a $\sigma_{v} \sim$200 km s$^{-1}$, shifted in flux for clarity. Vertical dotted lines mark the identified absorption features. Bottom: Fit residuals normalized to the error spectrum.}
     \label{fig:spec}
\end{figure}

To fully exploit the sample size we created an average spectrum. We scaled the wavelength vector of each galaxy to z$_{m}$ and interpolated their fluxes and error spectra on a 5\AA\ rest-frame grid, increasing the error in each new pixel by the square root of the width ratio between the old and the new spectral bin, to account for the introduced noise correlation. Each error spectrum was interpolated in quadrature. Each spectrum was normalized to its average flux between $\lambda_{rest}$=3800-3900 \AA\ in order to select a wavelength range covered for all galaxies. We then created a stacked spectrum as the inverse variance weighted mean of the individual fluxes in each wavelength bin. The final error spectrum was computed  as the error on the flux-weighted mean in each pixel. A jackknife resampling provides consistent results.
The median z$_{spec}$ uncertainty from our individual galaxies is dz$\sim$0.006 at 68\% confidence. This does not affect the final spectral resolution, being  three times smaller than the broadening from galaxy sizes. The average spectrum was rescaled to match the average broadband photometry of the sample at $\lambda_{rest}$=3800-3900 \AA, in turn obtained by stacking the individual galaxy best fit SEDs (as done for the spectra). Fig. \ref{fig:spec} shows the average spectrum with clear Balmer absorption lines from H$\eta$ to H$\gamma$, the [O II] (unresolved) doublet at 3727 \AA\ and the FeI absorption line. With a mean S/N $\sim$13 per pixel at 4000\AA\ this is, to date, one of the highest S/N spectra available for high-\textit{z} PEGs and the first spanning a large spectral range around key spectral breaks.

We analysed the stacked spectrum as in section 2, this time limiting the age grid to the age of the Universe at z$_{m}$=2.808 (2.3~Gyr) and using the average FWHM derived for the individual targets as spectral broadening.\footnote{This analysis does not include the information from the stacked photometry.}
The mean properties of our sample result in $t_{50}=0.30^{+0.20}_{-0.05}$ Gyr and Av=0.6$^{+0.6}_{-0.4}$ mag at fixed solar metallicity given the high $<$M$_{\star}>$ of our sample \citep{mancini19} (see Table \ref{tab:params}). 
The 1$\sigma$ uncertainties were estimated marginalizing over these two quantities, taking into account the simple parameterizations used for the SFHs.

\afterpage{
\begin{figure}[t!]
\centering
\includegraphics[width=\linewidth]{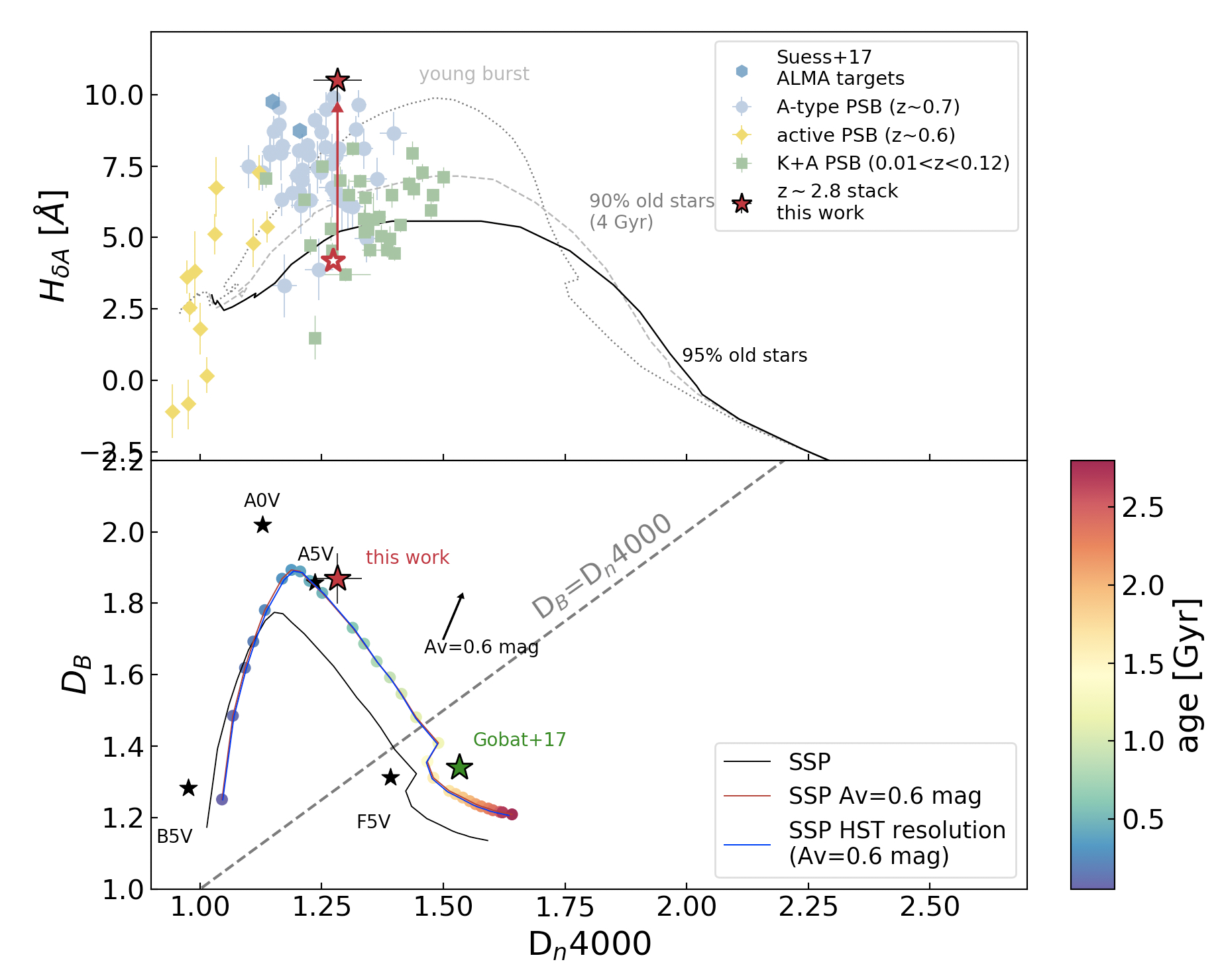}
\caption{Top: comparison between our average $D_{n}4000$ and $H_{\delta A}$ and z$\sim$0.7 A-type post-starburst galaxies (light blue dots) from \cite{suess17}. Green squares and yellow diamonds: K+A local post-starbursts and z$\sim$0.6 active post-starbursts, \cite{french15,sell14} respectively. Open star: H$\delta_{A}$ computed on the stacked spectrum, as a lower limit due to resolution. Tracks with different amounts of old stars (age$>$4Gyr) from \cite{suess17} are shown, going from no old stars (dotted line) to 95\% of old stars (solid line). Bottom: strength of the Balmer and 4000 \AA\ break indices as a function of the SSP age. Red and green stars represent the indices measured on our stack spectrum and on \cite{g17} spectrum respectively. Black stars mark the indices computed on theoretical stellar spectra smoothed to HST resolution \citep{uvblue}.} 
\label{fig:hdelta}
\end{figure}
}

\begin{figure*}
\centering
\includegraphics[width=\linewidth]{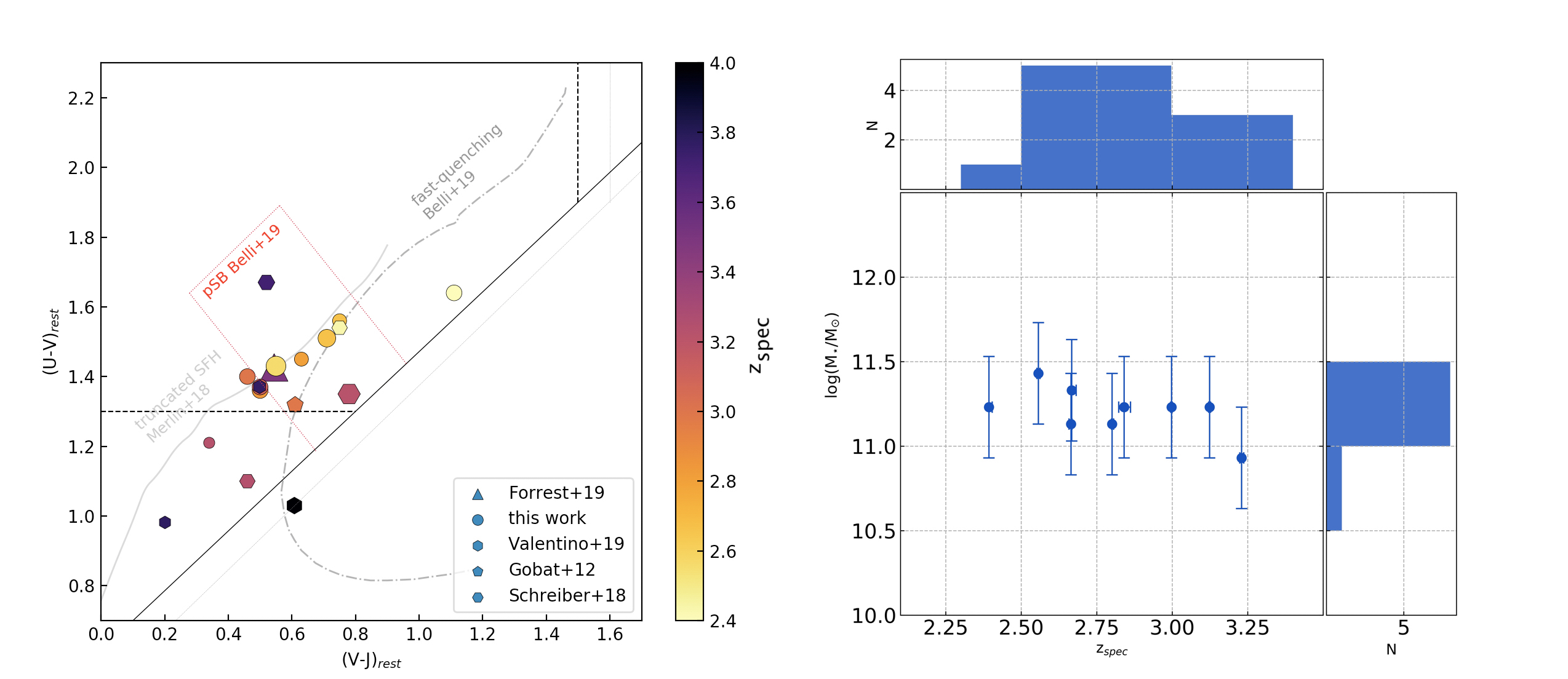}
\caption{Left: \textit{UVJ} colors of high-redshift ($z>2.5$) quiescent galaxies color coded for their spectroscopic redshift. Solid and dotted-dashed tracks show the color evolution for fast-evolving SFHs, from \cite{merlin18} and \cite{Belli19} respectively. The red box shows the post-starburst region as defined in \cite{Belli19}. Marker sizes are proportional to stellar mass. Right: stellar mass and redshift distributions of the sample.}
\label{fig:uvj}
\end{figure*}

We then measured the $D_{n}$4000, $D_{B}$ and $H_{\delta A}$ spectral indices \citep{worthey97,balogh99, kriek06}. Since our spectral resolution is worse than in the original Lick system, the H$_{\delta A}$ was measured on the best-fit BC03 model smoothed to the Lick resolution and adding the broadening due to a 200 km s$^{-1}$ stellar velocity dispersion corresponding to our $<$M$_{\star}>$. The $D_{n}$4000 and $D_{B}$ instead are not affected much by resolution. The indices result in $D_{n}$4000=1.28$\pm$0.05, $D_{B}$=1.88$\pm0.07$  and H$_{\delta A}$=10.1 $^{+0.0}_{-0.6}$\AA. The error on $H_{\delta A}$ was computed measuring the index maximum variation on the templates within the uncertainties on t50 and Av, accounting for the negative correlation between these two. The weakness of the D$_{n}$4000 compared the strength of the Balmer Break ($D_{B}$) and of $H_{\delta A}$ classifies the spectrum as post-starburst dominated by A-type stars. We show the indices in Fig.\ref{fig:hdelta}, where we also show H$_{\delta A}$ measured directly on the stacked spectrum to display the effect of low resolution. 
We checked whether our spectroscopic sample is representative of the parent photometric sample and quantified the impact of the selection bias introduced by the H-band selection at 2.5$\leqslant$z$\leqslant$3.5.
A Kolmogorov–Smirnov test reveals that the probability of being drawn from the same M/L distribution is only 3\%, confirming the selection bias. However, removing 30\% of the highest M/L objects from the parent sample yields $p=0.5$, leaving our HST sample consistent with the remaining 70\%. The median M/L ratios of the two samples become identical once the highest M/L objects, 40\% of the parent sample, are removed. The lowest M/L that a non post-starburst galaxy (defined here as a stellar population characterized by D$_{B}$/$D_{n}4000<$1) would have relative to that of our average galaxy implies a cut ($M/L_{old}\geqslant$ M/L$_{HST}~+~\Delta M/L$) that includes at most 30\% of the unobserved sample. 
We thus conservatively conclude that at least 60-70\% of passive galaxies at $z\sim3$ are post-starburst, as the remaining ''redder" galaxies could be as well post-starbursts with higher dust extinction, but also redshift interlopers, star forming interlopers, or AGN (our single failed confirmation from spectroscopy is among the reddest and lowest M/L galaxies in the parent sample). Hence, the typical passive galaxy at $z\sim3$ is a post-starburst.
This conclusion is corroborated by the global fast rise of the passive stellar mass function over $2.5<z<4$ \citep{muzzin13, iary17} consistent with a $\sim$1 dex increase of the number density, implying that a large proportion of the massive passive population in place at $z\sim3$ must have been very young and recently formed. In addition, from Fig.\ref{fig:uvj} and from the strong Balmer lines in the published spectra, it appears that all z$\geqslant$3 spectroscopically confirmed passive galaxies from the literature are sharing our global spectral properties. Further efforts need to be taken to spectroscopically identify truly old and passive galaxies (non post-starburst) at high redshift, if they exist.

%

\section{Residual star formation}

Given the tight anticorrelation between the age of a stellar population and the specific SFR (sSFR=SFR/$<M_{\star}>$ ) of its host galaxy, expected in the case of a smooth intrinsic SFH (as opposed to a simple stellar population (SSP) or a complex SFH), we investigated whether the age decrease at fixed M$_{\star}$ from z$\sim$1.5 to z$\sim$2.8 translates into a higher sSFR.

We used MPFIT to model the residuals of the fit using a single Gaussian centered and fixed at 3727 \AA\ with a width matched to the data spectral resolution, measuring F$_{[O II]}$=(3.1 $\pm$ 0.7) $\times$ 10$^{-18}$ erg s$^{-1}$ cm$^{-2}$.
The observed [O II] flux was dereddened adopting the best fit dust attenuation (A$_{V}$=0.6 mag) using a value of \textit{f}=0.83 for the ratio between stellar and nebular extinction \citep{kashino13} as in \cite{g17} in order to perform a consistent relative comparison. Following \cite{kenn98} we then obtained an SFR=$7\pm3$ $M_{\sun}$ yr$^{-1}$ which is a factor of $\sim$ 60 below the main sequence (MS) level at $2.5<z<3.5$ \citep{schr15}. This is in agreement with estimates for intermediate redshift quiescent galaxies from dereddened [O II] luminosity (SFR$_{[O~II]}$=4.5 M$_{\odot}$yr$^{-1}$, \cite{g17}) which is in turn consistent with their FIR luminosity (SFR$_{IR}$=$4.8^{+1.8}_{-1.3}$ M$_{\odot}$yr$^{-1}$, \cite{g18}, hereafter G18). Therefore the age evolution implied by the continuum appears to occur at constant sSFR. This is also suggested by the equivalent width (EW) of the line which is 2.1 $\pm$ 0.6 \AA\ rest-frame. This value is almost half that obtained by \cite{g17} due to a higher UV continuum produced by $\sim$1 Gyr younger stars.
Assuming that the [O II] flux arises from a 100 Myr-old stellar population, forming stars at a constant rate and with Av=0.6 mag, we subtracted the associated continuum component from the stacked spectrum to explore the effect on the age determination, deriving $t_{50}=0.30^{+0.20}_{-0.05}$ Gyr and Av=0.9$^{+0.4}_{-0.5}$ mag. Letting the age and extinction of the star-forming component vary (100-300 Myr and 0.6-1.5 mag, respectively) leaves t$_{50}$ unchanged while increasing Av up to 1.0 mag. The flux at $\lambda >$4400 \AA\ varies around the 95\% of the original flux, suggesting that the CSF component does not account for more than 1-2\% of $<$M$_{\star}>$ as also confirmed by the mass produced in a $\Delta$t $\sim$300 Myr at a constant SFR of 10 M$\sun$ yr$^{-1}$, namely the upper limit inferred here. We thus conclude that the presence of the youngest stellar population does not impact our results significantly.
The sSFR of our sample, (4.35 $\pm$ 2.47) $\times10^{-11}$ yr$^{-1}$ is consistent with the available estimates at z$\sim$1.5-1.8 \citep{ sar15, bez19} and with the sSFR for the 4 quiescent galaxies with a spectral break in \cite{schr18b} for which sSFR$_{[O~II]}$ is available. However, the best fit t$_{50}$ here obtained implies a formation redshift z$_{form} \gtrsim 3.2$, meaning that at z$\sim$4 our galaxies would be actively forming stars. 
To check that the average spectrum is not dominated by emission from unobscured AGN, we removed from the stack the two X-ray detected sources (L$_{X_{2-10keV}}=10^{44.3}$ and $ 10^{43.7}$ erg  s$^{-1}$), getting F$_{[O~II]_{noX}}$ =(3.53 $\pm$ 0.90) $\times$ 10$^{-18}$ erg s$^{-1}$cm$^{-2}$ consistent within the uncertainties. Stacking the available X-ray data in COSMOS for the undetected sources results in L$_{X_{2-10keV}}<$10$^{43.1}$ erg s$^{-1}$. 
If part of the [O II] luminosity were caused by low AGN activity or shocks, the intrinsic SFR would be even lower, strengthening the conclusion that the young spectral age does not directly map into a higher sSFR.
Using the UV-extended \cite{m09} models the best-fitting template yields a similar t$_{50}$ $\sim$0.3 Gyr but produces a slightly worse $\chi^{2}$ and a less solid [O II] detection, with a flux that is 60\% the one measured with BC03 templates, at 2.9 $\sigma$ confidence.\\

\section{Discussion} 

It is interesting to place our $z\sim3$ results in the overall evolutionary context of passive galaxies. In order to be able to compare to the available literature we express, for convenience, the SFR constraints in terms of the available gas fraction through the relation $f_{mol}=sSFR/SFE$, where SFE is the star formation efficiency. We caution that this is just an alternative way to look at the SFR result, as we are using the same single constraint: we use both quantities interchangeably in the following and notably in Fig.4.
We assume the same SFE derived in G18 ($5\times10^{-10}$ yr$^{-1}$), which is lower by $\times2$--3 than that of typical star forming galaxies, noticing that such reduced SFE is also typical of post-starburst galaxies \citep{suess17}.
Our SFR thus converts into $M_{mol}$=(1.5$\pm$0.6)$\times$10$^{10}$M$_{\odot}$, hence f$_{mol}\sim9\pm4$\%.
We compare it to CO or dust-continuum-based gas fractions and upper limits (converted to Salpeter) for quiescent and post-starburst galaxies: \cite{davis14} and \cite{saint11} for local massive PEGs; \cite{sar15}, \cite{bez19}, \cite{spilk18}, \cite{zavala19}, \cite{rud17a}, \cite{suess17}, \cite{spilk18}, \cite{hayashi18}, \cite{g18} for intermediate-\textit{z} quiescent galaxies; \cite{schr18b} and \cite{Valentino19} for z$\sim$3--4 galaxies.
Despite the uncertainties, our data at z$\sim$3 seems to disfavor the steep (1+z)$^{4-5}$ trend inferred from $z=0$ to 1.5--2, suggesting a flattening in the M$_{mol}$/M$_{\star}$ evolution (or equivalently, of the sSFR).\\
The published SFR$_{[O~II]}$ for $z>3$ galaxies with a clear spectral break in \cite{schr18b} and \cite{Valentino19} also seem to support this trend.\\ 

Our data thus suggest that a substantial fraction of the massive quiescent population at z$\sim$3 is approaching the quenching epoch, becoming intrinsically young and with spectra of post-starburst galaxies, but with a more or less constant f$_{mol}$ (or equivalently, sSFR) over $1.5<z<3$--4. Various processes could be simultaneously acting: cosmological cold flows, AGN feedback, dust grain growth in the interstellar medium, satellite accretion or dust destruction by sputtering with a hot X-ray halo. It is beyond the scope of this Letter to investigate this further.

\begin{figure}
\centering 
\includegraphics[width=\columnwidth]{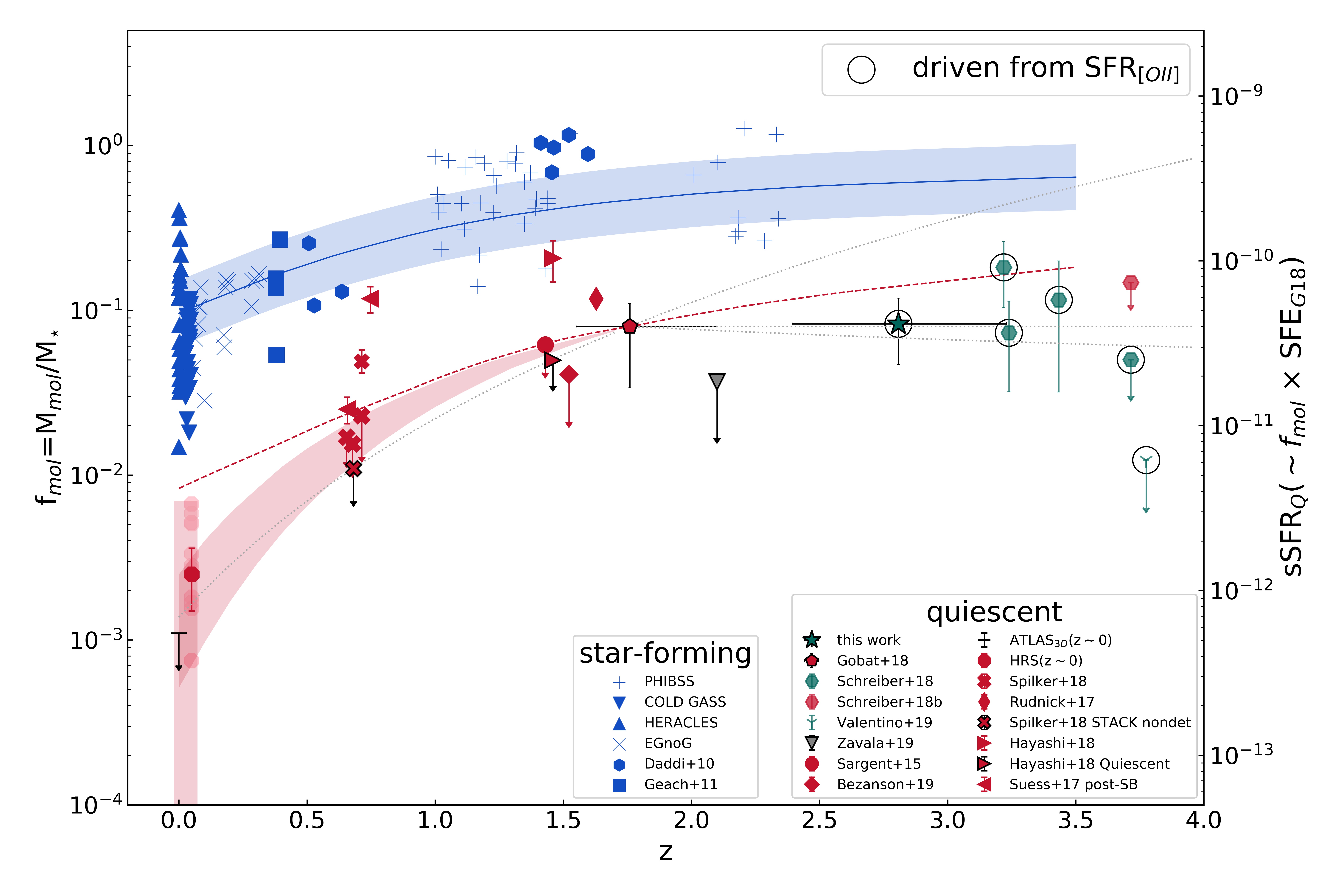}
  \caption{Evolution of f$_{mol}$ or equivalently (see Sect. 5) the sSFR of quiescent galaxies (sSFR$_{Q}$) with redshift, adapted from \cite{g18}. Red symbols mark quiescent galaxies, old and post-starbursts, with CO or dust-continuum measures, whereas circled green symbols mark estimates from SFR$_{[O~II]}$ alone, converted to f$_{mol}$ adopting G18 SFE. Black contours mark stacked samples, where horizontal error bars show the spread in redshift of individual targets. Blue symbols mark the gas fraction of low-, intermediate-, and high-\textit{z} ($z>1$) MS galaxies. The blue curve and shaded area mark the evolution of an average 5$\times10^{10}$ M$_{\sun}$ MS galaxy and its 0.2 dex scatter \citep{sar14}. The red dotted curve shows the same relation, offset by a factor of 6 and for a stellar mass of log($M^{*}(z)$+$\Delta$M), where $\Delta$M is the offset between the median mass of G18 sample and the M$^{*}$ of the passive stellar mass function at $<$z$>\sim$1.8. The pink shaded area shows the trend derived in G18. Grey lines are extrapolations of the low redshift trend assuming a gas fraction rise as (1+z)$^{\alpha}$, with $\alpha \sim$4-5, as fast as in G18; $\alpha$=2.2 as for MS galaxies; $\alpha$=0 for no evolution; and $\alpha$=-0.5 representative of a negative evolution.}
\label{fig:fmol}
\end{figure}

\begin{table*}
\centering
\caption{ Main properties of our sample derived from the stacked spectrum (see text). Solar metallicity and a Salpeter IMF were assumed throughout.}
\label{tab:params}
\renewcommand\arraystretch{1.4}
\begin{tabular}{cccccccccc}
\hline
$<$z$>$ & $<M_{\star}>$ & $t_{50}$ & Av & D$_{n}$4000 & D$_{B}$ & EW$_{[O~II]}$ & F$_{[O~II]}$ & SFR \\

 & ($10^{11}$ M$_{\odot}$) & (Gyr) & (mag) & & &  (\AA ) & (erg s$^{-1}$  cm$^{-2}$) &(M$_{\odot}$ yr$^{-1}$)\\
\hline
\hline
2.808  & 1.8 $\pm$ 0.8 & $0.30^{+0.20}_{-0.05}$  & 0.9$^{+0.4}_{-0.5}$ & 1.28 $\pm$ 0.05 & 1.88$\pm$0.07 & 2.1 $\pm$ 0.6 & 3.1$\pm$ 0.7 $\times 10^{-18}$& 7 $\pm$ 3 \\

\hline 
\end{tabular}
\end{table*}

One could wonder if highly obscured star formation could be present and go unrecognized, given that their post-starburst nature might suggest that a highly obscured starburst might have been previously present.
Stacking at 3GHz (excluding two clear radio AGN detections at S$_{3GHz}$ 0.58$\pm$0.03~mJy and 14$\pm$4 $\mu$Jy) results in a $3\sigma$ signal with S$_{3GHz}$=2.72$\pm$0.93 $\mu$Jy which translates into a rest-frame L(1.4~GHz)~$\sim$ 2$\times$10$^{23}$~W/Hz,  4 times higher than in \cite{g18}. 
This is probably just suggesting a higher radio AGN activity at fixed stellar mass closer to the quenching epoch, continuing the rapid evolution seen from $z=0$ to 1.5--2. 
On the other hand, such a detection would also be formally consistent with  $40-50~M _{\odot}$ yr$^{-1}$ of obscured star formation (assuming the IR--radio correlation at $z=3$, \cite{FRC}). This would still place the typical object 8 times below the MS. While we tend to interpret strong Balmer absorption lines plus weak [O II] emission as a sign of post-starburst galaxies with residual unobscured star formation, we cannot fully rule out stronger obscured SF, which could be analogous to what seen locally for \textit{e(a)} galaxies \citep{poggianti2000}. Only future \textit{Atacama Large Millimeter/submillimeter Array} observations of several of these targets could conclusively solve this issue. 

\begin{acknowledgements} 
We thank the anonymous referee for their constructive report and Gabriel Brammer for his help with \texttt{grizli}. A.C. acknowledges grants PRIN MIUR 2015, PRIN 2017 (20173ML3WW\_001), and ASI n.I/023/12/0. S.J. acknowledges MICIU grant AYA2017-84061-P, co-financed by FEDER. I.D. is supported by 
Marie Sk\l{}odowska-Curie grant agreement No. 788679.\\ 

\end{acknowledgements}

\newpage
%
%

%
%

\end{document}